\documentclass[aps,twocolumn,showpacs,floats,prl]{revtex4}
\usepackage{graphicx}
\usepackage{dcolumn}
\usepackage{bm}
\usepackage{color}

\begin{document}

\author{N. Matveeva}
\author{S. Giorgini}
\affiliation{Dipartimento di Fisica, Universit\`a di Trento and CNR-INO BEC Center, I-38050 Povo, Trento, Italy}

\title{Liquid and crystal phase of dipolar fermions in two dimensions} 

\begin{abstract} 
The liquid and crystal phase of a single-component Fermi gas with dipolar interactions are investigated using quantum Monte Carlo methods in two spatial dimensions and at zero temperature. The dipoles are oriented by an external field perpendicular to the plane of motion, resulting in a purely repulsive $1/r^3$ interaction. In the liquid phase we calculate the equation of state as a function of the interaction strength and other relevant properties characterizing the Fermi-liquid behavior: effective mass, discontinuity at the Fermi surface and pair correlation function. In the high density regime we calculate the equation of state of the Wigner crystal phase and the critical density of the liquid to solid first order phase transition. Close to the freezing density we also search for the existence of a stripe phase, but such a phase is never found to be energetically favorable.  
\end{abstract}

\maketitle 

The recent rapid developments in the field of ultracold dipolar atoms and molecules has opened up new fascinating prospects for investigating many-body effects in quantum degenerate gases with long-range interactions (for a review see, {\it e.g.}, Ref.~\cite{Review-Baranov}). In this respect, single-layer and multi-layer configurations of two-dimensional fermions are particularly intriguing because of the competing interplay, depending on the strength of the dipolar interaction and on the distance between layers, between Fermi liquid behavior, superfluid pairing, crystal order and density-wave instabilities~\cite{Bruun, Yamaguchi, DasSarma, Santos, Zinner, Demler, Baranov, Parish, Shlyapnikov}. 

Fermionic molecules of $^{40}$K$^{87}$Rb, which can have a strong electric dipole moment, have been created using coherent transfer of Feshbach molecules to their rovibrational ground state~\cite{Jila1} and have been brought toward the quantum degenerate regime~\cite{Jila2}. Other fermionic molecules are now being actively studied experimentally~\cite{Zwierlein, Ketterle}. Atomic species with a large magnetic moment, such as dysprosium, offer a different possibility of realizing degenerate Fermi gases of dipoles that was successfully pursued in the recent experiment reported in Ref.~\cite{Lev}. 

A particularly simple geometrical arrangement of a single-component dipolar Fermi gas in 2D is when the dipoles are oriented perpendicular to the plane of motion by means of a sufficiently strong external field. This configuration has been proven to greatly suppress the chemical reaction rate of molecules, thereby enhancing their lifetime~\cite{Jila3}. Here particles at distance $r$ interact via a purely repulsive, rotationally symmetric and long range $1/r^3$ potential. Still the phase diagram at zero temperature is expected to be quite rich: interlayer dimers and a novel BCS-BEC superfluid crossover are predicted in bilayer systems~\cite{Santos}, while in-plane and out-of-plane density ordered phases are predicted in multilayer systems~\cite{Zinner, Demler}. In the case of a single layer, a Fermi liquid with peculiar scattering properties is stable at low density~\cite{Shlyapnikov} and a Wigner crystal emerges at high density, where the classical potential energy of dipoles largely exceeds their kinetic energy. For intermediate values of the interaction strength an instability at finite wave vector is predicted to set in~\cite{Yamaguchi, DasSarma, Parish}, driving the system to a stripe phase that breaks both rotational and translational symmetry (in the direction perpendicular to the stripes). A similar scenario, involving microemulsion phases ({\it e.g.} stripes or bubbles) is expected for the meting of the Wigner crystal at $T=0$ in a 2D Coulomb gas\cite{Spivak}. These results are derived within a mean-field approach: an important question concerns the quantitative determination of the phase diagram using more accurate theoretical tools, such as quantum Monte Carlo (QMC) techniques~\cite{Note1}.  

\begin{figure}
\begin{center}
\includegraphics[width=7.5cm]{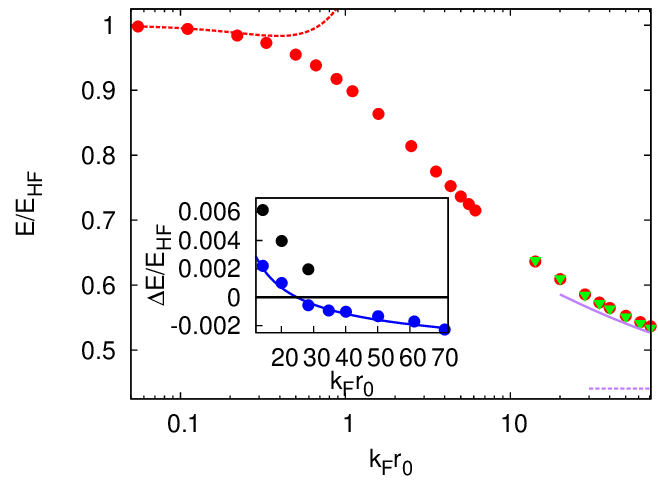}
\caption{(color online). Equation of state of the liquid and solid phase in units of the Hartree-Fock energy (\ref{HF}). Circles refer to the liquid and triangles to the solid. The red dashed line corresponds to the second-order expansion in Ref.~\cite{Shlyapnikov}. The purple dashed horizontal and solid line correspond respectively to the classical energy of the Wigner crystal and to the result of Ref.~\cite{Mora} including the first correction arising from the zero-point motion of phonons. Inset: Energy difference between the solid and the liquid (blue circles) and between the stripe phase and the liquid (black circles). The blue solid line is obtained from a best fit to the equation of state of the liquid and solid phase. Error bars are smaller than the symbols size.}
\label{fig1}
\end{center}
\end{figure}

In this Letter we examine a 2D system of $N$ identical fermionic particles of mass $m$ that interact with the Hamiltonian
\begin{equation}
H=-\frac{\hbar^2}{2m}\sum_i\nabla_i^2+\sum_{i<j}\frac{d^2}{r_{ij}^3} \;,
\label{Hamilton}
\end{equation}
where $r_{ij}$ is the distance between particle $i$ and $j$ and $d$ is the intensity of the electric (or magnetic) dipole moment. 
The strength of the dipolar interaction is conveniently expressed in terms of the dimensionless parameter $k_Fr_0$, where $r_0=md^2/\hbar^2$ is the characteristic length of the dipole-dipole force and $k_F=\sqrt{4\pi n}$ is the Fermi wave vector of the 2D gas determined by the density $n$. The energy scale set by $k_F$ is the Fermi energy $\epsilon_F=\hbar^2k_F^2/2m$. As a  function of $k_Fr_0$ we investigate the ground-state properties of the Fermi liquid phase including the equation of state, the effective mass, the discontinuity of the momentum distribution at $k_F$, the pair correlation function and the static structure factor. By comparing the energy of the Fermi liquid (FL) and of the Wigner crystal (WC) phase, we determine the value $k_Fr_0=25\pm3$ of the freezing density. Furthermore, in the region of interaction strengths close to freezing, we calculate the energy corresponding to a stripe phase finding that it is never favorable compared to the FL and WC state. The main results on the equation of state are shown in Fig.~\ref{fig1} in units of the Hartree-Fock energy
\begin{equation}
E_{HF}=N\frac{\epsilon_F}{2}\left(1+\frac{128}{45\pi}k_Fr_0 \right) \;,
\label{HF}
\end{equation} 
corresponding to the lowest order perturbation expansion of the FL in the interaction parameter $k_Fr_0$\cite{Yamaguchi, Shlyapnikov}.

We use the fixed-node diffusion Monte Carlo method (FN-DMC), a projector technique that, starting from an antisymmetric trial wave function $\psi_T$, finds the state having the lowest energy compatible with the many-body nodal surface of $\psi_T$ that is kept fixed during the calculation. The method provides a rigorous upper bound to the energy of the fermionic ground-state~\cite{FNDMC}. 

Simulations are performed in a box of volume $V=L_xL_y$, where we always take $L_x\le L_y$. The density is $n=N/V$ and we use periodic boundary conditions (PBC) in both spatial directions. Since the dipole-dipole interaction is long range, the potential energy contribution to the Hamiltonian, given by the second term in Eq.~(\ref{Hamilton}), requires a careful treatment
\begin{equation}
V_{dd}=\sum_{i<j}\frac{d^2}{|{\bf r}_i-{\bf r}_j|^3} + \frac{1}{2}\sum_{i,j}\sum_{{\bf R}\neq0} \frac{d^2}{|{\bf r}_i-{\bf r}_j-{\bf R}|^3} \;,
\label{Hamilton1}
\end{equation} 
where $i$ and $j$ label particles in the simulation cell and the vectors ${\bf r}_{i(j)}+{\bf R}$ correspond to the positions of all images of particle $i(j)$ in the array of replicas of the simulation cell. The combination of all images has the same average density $n$ of the simulation box and provides a good approximation of the homogeneous medium. In the case of the Coulomb potential the summation in (\ref{Hamilton1}) is carried out by means of the Ewald method~\cite{FNDMC}. For the faster convergent $1/r^3$ potential the mean interaction energy can be evaluated using the simpler formula
\begin{equation}
\langle V\rangle=(V_{dd})_{R_c}+V_{\text{tail}} \;,
\label{Hamilton2}
\end{equation}
where $(V_{dd})_{R_c}$ denotes the sum (\ref{Hamilton1}) with the constraint $|{\bf r}_i-{\bf r}_j-{\bf R}|\le R_c$ and $V_{\text{tail}}=\pi n d^2/R_c$ is the contribution from distances larger than $R_c$ assuming a uniform distribution of particles. The cut off length $R_c$ is chosen large enough ($R_c=2L_x$) to yield an average interaction energy $\langle V\rangle$ that is independent on $R_c$, within statistical uncertainty.

\begin{figure}
\begin{center}
\includegraphics[width=6.5cm]{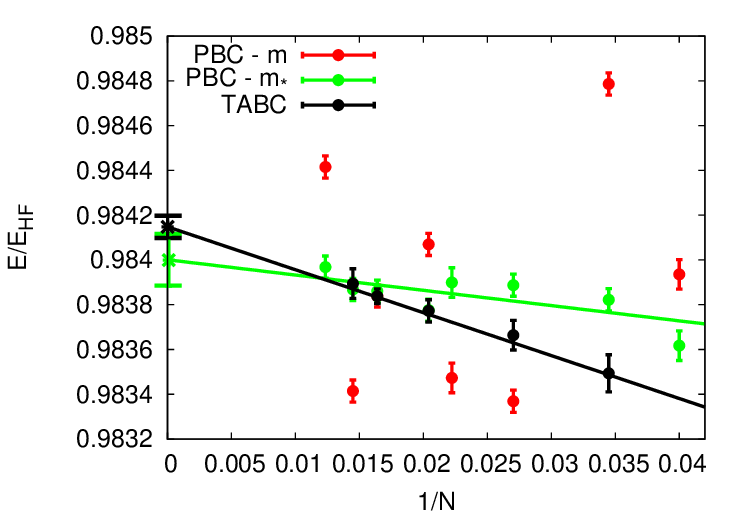}
\caption{(color online). Finite size effects in a Fermi liquid at the density $k_Fr_0=0.22$. Black symbols and line correspond to TABC energies and linear fit. Green symbols and line correspond to PBC energies $E_N-(m/m^\ast)\Delta T_N$ and linear fit. Red symbols correspond instead to PBC energies $E_N-\Delta T_N$. The values extrapolated to the TL are shown with the corresponding error bars.}
\label{fig2}
\end{center}
\end{figure}

\underline{\it{Fermi liquid phase}:} The trial wave function describing the FL phase is assumed of the Jastrow-Slater form
\begin{equation}
\psi_T({\bf r}_1,...,{\bf r}_N)=\prod_{i<j}f(r_{ij}) \; \det[e^{i{\bf k}_\alpha\cdot{\bf r}_i}] \;.
\label{FL1}
\end{equation}
Here ${\bf k}_\alpha=(2\pi/L)(n_\alpha^x,n_\alpha^y)$ with $n_\alpha^{x,y}=0,\pm1,\pm2,...$ are the wave vectors complying with PBC in the square box $(L_x=L_y=L)$ and $f(r)$ is a non-negative function satisfying the boundary condition $f^\prime(r=L/2)=0$. The short-range behavior of $f(r)$ is of the form $f(r)\propto K_0(2\sqrt{r_0/r})$, where $K_0$ is the modified Bessel function, and fulfills the cusp condition of the dipole-dipole potential~\cite{Astra}. 

A delicate issue related to QMC calculations of the equation of state is the extrapolation to the thermodynamic limit (TL). In the FL phase apart from the size dependence affecting the potential energy contribution, which we treat using the procedure in Eq.~\ref{Hamilton2}, significative shell effects are present in the kinetic energy contribution. We consider closed-shell configurations corresponding to $25\le N\le81$ for which the relative error $|\Delta T_N|/E_{TL}^{(0)}=|E_N^{(0)}/E_{TL}^{(0)}-1|$ in the energy of the non-interacting gas compared to the TL can be as large as $\sim1\%$. An extrapolation method based on FL theory is provided by the fitting formula~\cite{Ceperley1}
\begin{equation}
E_N=E_{TL}+\frac{m}{m^\ast}\Delta T_N+\frac{a}{N} \;,
\label{extrapolation}
\end{equation}  
which involves the parameter $m/m^\ast$, determining the inverse effective mass of the particles, and the coefficient $a$ of the residual size dependence assumed to be linear in $1/N$. Here $E_N$ is the QMC output energy of the $N$-particle system with the potential contribution evaluated using Eq.~\ref{Hamilton2}.  The values of $E_N-\Delta T_N$ are shown as red symbols in Fig.~\ref{fig2}. Their scattered dependence on $1/N$ is considerably suppressed if one accounts for the effective mass, as it is shown by the green symbols corresponding to $E_N-(m/m*)\Delta T_N$. A more reliable convergence to the TL is obtained using the method of twist-averaged boundary conditions (TABC)~\cite{Ceperley2}. Here the PBC wave vectors of the plane waves in the Slater determinant of Eq.~(\ref{FL1}) are replaced by ${\bf k}_\alpha(\boldsymbol{\theta})=(2\pi/L)(n_\alpha^x+\theta_x,n_\alpha^y+\theta_y)$, where $\theta_x$, $\theta_y$ are continuous variables in the interval $[0,1]$. In the grand canonical implementation of TABC described in Refs.~\cite{Ceperley2, Chiesa} the wave vectors are constrained by $|{\bf k}_\alpha(\boldsymbol{\theta})|<k_F$ and different values of the twist $\boldsymbol{\theta}$ can correspond to different numbers of particles. The number of particles $\bar{N}$ and the energy $E_{\bar{N}}$ are obtained from averages over all possible twist angles. With our use of TABC we still find a residual size effect $\Delta T_{\bar{N}}$~\cite{Note2}. The extrapolation to the TL is performed using Eq.~(\ref{extrapolation}) and  statistical agreement in $E_{TL}$ between PBC and TABC is obtained for all values of the density (see Fig.~\ref{fig2}).

The FN-DMC results of the ground-state energy are reported in Fig.~\ref{fig1}. At low density we find good agreement with the result $E=E_{HF}+(N\epsilon_F/2)(k_Fr_0)^2\log(1.43k_Fr_0)$, which was  derived in Ref.~\cite{Shlyapnikov} including corrections to the lowest order expansion (\ref{HF}). 

\begin{figure}
\begin{center}
\includegraphics[width=6.5cm]{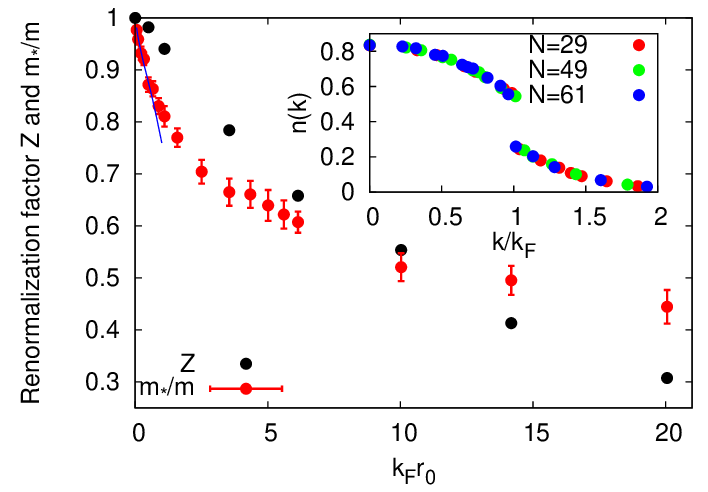}
\caption{(color online). Effective mass and renormalization factor in the liquid phase as a function of the interaction strength. The line corresponds to the perturbation expansion for $m^\ast/m$ of Ref.~\cite{Shlyapnikov}. Inset: Momentum distribution corresponding to $k_Fr_0=20$ for different system sizes.}
\label{fig3}
\end{center}
\end{figure}

\begin{figure}
\begin{center}
\includegraphics[width=6.5cm]{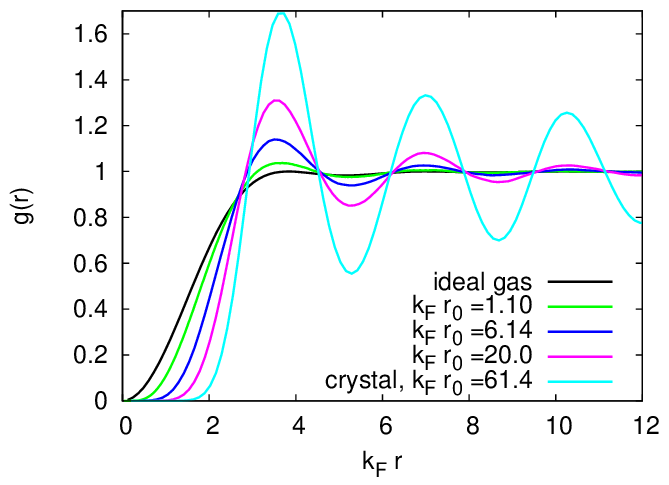}
\caption{(color online). Pair correlation function in the liquid and in the crystal phase. The pair correlation function of the non-interacting gas is also shown.}
\label{fig4}
\end{center}
\end{figure}

\begin{figure}
\begin{center}
\includegraphics[width=6.5cm]{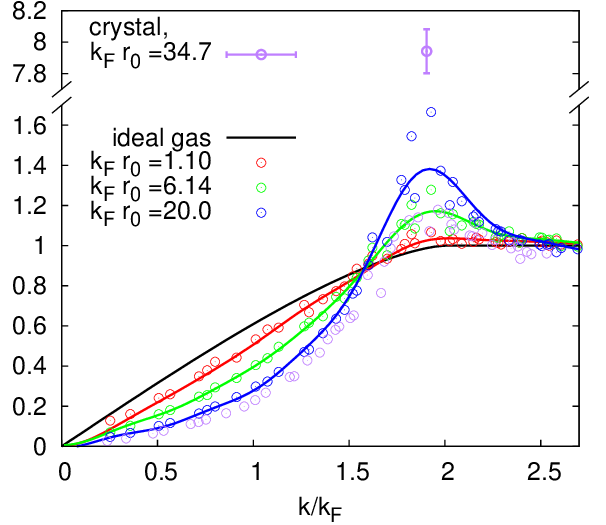}
\caption{(color online). Static structure factor in the liquid and in the crystal phase. In the liquid phase, solid lines correspond to the Fourier transform of $g(r)$ while symbols correspond to the direct calculation of $S(k)$. The static structure factor of the non-interacting gas is also shown.}
\label{fig5}
\end{center}
\end{figure}

The effective mass $m^\ast$ obtained from Eq.~(\ref{extrapolation}) is shown in Fig.~\ref{fig3} for different values of $k_Fr_0$. At weak coupling our results well reproduce the perturbation expansion reported in Ref.~\cite{Shlyapnikov}, but for larger couplings the reduction of $m^\ast$ is less pronounced than the perturbative prediction and $m^\ast/m$ approaches the value 0.4 for $k_Fr_0$ close to freezing. We also calculate the momentum distribution $n(k)$ and using the fit  $n_k=Z\theta(k_F-k)+g(k)$, where $\theta(x)$ is the step function and $g(k)$ is a continuous function of $k$, we extract the renormalization factor $Z$. Results are reported in Fig.~\ref{fig3}. In the inset of Fig.~\ref{fig3} we show $n_k$ in the strongly interacting regime for different numbers of particles. It is worth noticing that finite size effects on this quantity are much less severe here than in the 2D Coulomb gas~\cite{Holzmann} and are similar to the case of hard-disks investigated in Ref.~\cite{Drummond}.

We finally calculate the pair correlation function $g(r)$ giving the probability of finding two particles at the distance $r$. Results for different values of $k_Fr_0$ in the FL phase are shown in Fig.~\ref{fig4}. It is interesting to notice that by increasing the interaction strength the short-range repulsion increases and a shell structure starts to appear on approaching the freezing density. The Fourier transform of $g(r)$ yields the static structure factor $S({\bf k})=1+n\int d{\bf r} e^{i{\bf k}\cdot{\bf r}}[g(r)-1]$. This quantity can also be calculated directly in the FN-DMC algorithm by evaluating the average of the product of density fluctuation operators $NS({\bf k})=\langle\rho_{\bf k}\rho_{-{\bf k}}\rangle=\langle\sum_{i,j}e^{i{\bf k}\cdot({\bf r}_i-{\bf r}_j)}\rangle$. Results are reported in Fig.~\ref{fig5} for both estimators~\cite{Note3}. For large values of $k_Fr_0$ the direct estimator exhibits a more pronounced peak compared to the smoother Fourier transform at the wave vector corresponding to the lowest non-zero reciprocal lattice vector of the triangular lattice.

\underline{{\it Wigner crystal phase}:} To describe the WC phase we make use of the following trial wave function
\begin{equation}
\psi_T({\bf r}_1,...,{\bf r}_N)=\prod_{i<j}f(r_{ij}) \; \det[e^{-({\bf r}_i-{\bf R}_m)^2/\alpha^2}] \;,
\label{WC1}
\end{equation}
where the Jastrow correlation term $f(r)$ is the same as in the FL phase and the single-particle orbitals in the determinant are constructed with Gaussians, whose width $\alpha$ is a variational parameter, centered at the lattice points ${\bf R}_m$ of the triangular Bravais lattice. In order to enforce PBC, both the number of particles $N$ and the box sizes $L_x$ and $L_y$ must be multiples of a primitive cell, which can be chosen as a rectangle of side lengths $\ell_y=\sqrt{3}\ell_x$ containing two atoms. Extrapolation to the thermodynamic limit is obtained using a linear fit in $1/N$ over the FN-DMC energies. The results for the WC equation of state are reported in Fig.~\ref{fig1}. It is worth noticing that the antisymmetric constraint imposed in the wave function (\ref{WC1}) for particle exchange is negligible for the value of the energy. In fact, statistically compatible results are obtained with a node less wave function of distinguishable boltzmanons in agreement with the findings of Ref.~\cite{Astra}. This behavior is expected at large density, where the energy of the WC phase is given by the result~\cite{Mora}
\begin{equation}
E_{WC}=N\frac{\epsilon_F}{2}\frac{k_Fr_0}{4}\left( 1.597+\frac{2.871}{\sqrt{k_Fr_0}}\right) \;,
\label{WC2}
\end{equation}
obtained by including the contribution from the zero-point motion of phonons to the purely classical interaction energy of a system of dipoles arranged in a triangular Bravais lattice. The above expansion, holding for large $k_Fr_0$, is shown in Fig.~\ref{fig1} and is indeed approached by our QMC results. The difference between the ground state energy of the WC and FL phase is shown in the inset of Fig.~\ref{fig1}. From a fit to the equation of state of the two phases we can determine the intersection point at $k_Fr_0=25\pm 3$. This value is almost a factor two smaller compared to the critical density $k_Fr_0\sim 60$~\cite{Astra, Pupillo, Mora} of an equivalent system of bosons having the same mass, density and dipolar strength. This can be understood if one considers that the equation of state of the crystal is practically independent of statistics while the energy of the fermionic fluid is significantly larger than the bosonic one. From the equation of state of the FL and WC phase in the vicinity of the freezing density one can also estimate the width of the region where phase separation occurs driving the first-order liquid to solid transition. By imposing equilibrium of pressure and chemical potential in the two phases, the coexistence region turns out to be $\delta(k_Fr_0)\sim0.01$, a very small value consistent with a similar finding in the bosonic case~\cite{Mora}. The pair correlation function and the static structure factor deep in the crystal phase are shown respectively in Fig.~\ref{fig4} and Fig.~\ref{fig5}. In particular, $S({\bf k})$ exhibits a large peak at $k=1.90k_F$ corresponding to the lowest non-zero wave vector of the reciprocal lattice.
  
\underline{{\it Stripe phase}:} A pattern of equally spaced stripes is assumed in the $y$-direction corresponding to the trial wave function
\begin{equation}
\psi_T({\bf r}_1,...,{\bf r}_N)=\prod_{i<j}f(r_{ij}) \; \det[e^{ik_{\alpha x}x_i-(y_i-y_a)^2/\gamma^2}] \;,
\label{Stripes}
\end{equation}
where $y_a$ denotes the $y$ coordinate of the $a$-th stripe and $k_{\alpha x}=2\pi n_{\alpha x}/L_x$ are the PBC wave vectors in the $x$-direction. The number of fermions is the same in each stripe and once multiplied by the number of stripes determines the overall density in the volume $V=L_xL_y$. First, using the variational method, we optimize  the width $\gamma$ of the stripes and their separation $\Delta y=|y_{a+1}-y_a|$. For the latter quantity we find that $k_F\Delta y=\sqrt{4\pi}$ is an optimal value corresponding to a square box having $L_x=L_y$. We then perform FN-DMC simulations using PBC with different numbers of particles $N$ and we extrapolate to the thermodynamic limit relying on a $1/N$ linear fit. The results are shown in the inset of Fig.~\ref{fig1}, where we report the energy difference between the stripe and FL phase. For all values of $k_Fr_0$ in the vicinity of the freezing density the stripe phase is never energetically favorable compared to neither the FL nor the WC phase.

In conclusion, we investigated the ground state of a purely repulsive system of dipolar fermions and its liquid to solid transition. Important extensions of this work concern the effect of a tilting angle, making the interaction in the 2D plane anisotropic, and the coupling to a second layer inducing interlayer attraction. 

Useful discussions with G. Astrakharchik, M. Holzmann and M. Capone are gratefully aknowledged. This work has been
supported by ERC through the QGBE grant. Calculations were performed on the AURORA supercomputer at Fondazione Bruno Kessler.

\end{document}